\documentclass[prb,twocolumn,showpacs,preprintnumbers,amsmath,amssymb,a4paper,superscriptaddress]{revtex4-1}

\usepackage{graphicx}
\usepackage{dcolumn}
\usepackage{bm}
\usepackage{color}
\usepackage{amsmath}
\usepackage{amssymb}

\usepackage{sidecap}

\begin{document}

\title{Magnetic field-induced control of transport \\ in multiterminal focusing quantum billiards}

\author{C. Morfonios}
\affiliation{Zentrum f\"{u}r Optische Quantentechnologien, Universit\"{a}t Hamburg, Luruper Chaussee 149, 22761 Hamburg, Germany}
\author{D. Buchholz}
\affiliation{Theoretische Chemie, Institut f\"{u}r Physikalische Chemie, Universit\"{a}t Heidelberg, Im Neuenheimer Feld 229, 69120 Heidelberg, Germany}
\author{P. Schmelcher}
\affiliation{Zentrum f\"{u}r Optische Quantentechnologien, Universit\"{a}t Hamburg, Luruper Chaussee 149, 22761 Hamburg, Germany}

\date{\today}

\begin{abstract}
By exploring the four-terminal transmission of a semi-elliptic open quantum billiard in dependence of its geometry and an applied magnetic field, it is shown that a controllable switching of currents between the four terminals can be obtained.
Depending on the eccentricity of the semi-ellipse and the width and placement of the leads, high transmittivity at zero magnetic field is reached either through states guided along the curved boundary or focused onto the straight boundary of the billiard.
For small eccentricity, attachment of leads at the ellipse foci can yield optimized corresponding transmission, while departures from this behavior demonstrate the inapplicability of solely classical considerations in the deep quantum regime.
The geometrically determined transmission is altered by the phase-modulating and deflecting effect of the magnetic field, which switches the pairs of leads connected by high transmittivity.
It is shown that the elliptic boundary is responsible for these very special transport properties.
At higher field strengths edge states form and the multiterminal transmission coefficients are determined by the topology of the billiard.
The combination of magnetotransport with geometrically optimized transmission behavior leads to an efficient control of the current through the multiterminal structure.
\end{abstract}


\maketitle

\section{Introduction}

Open particle billiards are two-dimensional (2D) scattering structures that confine ballistically moving (charged) particles within a region of space of certain geometry, at the boundary of which openings allow the particle to escape into leads.
Their quantum version models nanoscale transport devices, usually realized by controllably confining a 2D electron gas at a semiconductor interface, and sets the grounds for the theoretical description and experimental investigation of coherent transport in the mesoscopic regime.
The transport properties of such billiards are extensively examined in the context of various interesting phenomena, including the quantization of conductance in mesoscopic systems,\cite{Picciotto2000,Chen2009}
the appearance of Fano resonances in transmission,\cite{Clerk2001,Mendoza2008,Baernthaler2010,Miroshnichenko2010} resonance trapping,\cite{IRotter2009b,Racec2010} shot noise in quantum dots,\cite{Blanter2000,Aigner2005} Andreev tunneling and reflection,\cite{Fazio1997,Fytas2005} as well as conductance fluctuations,\cite{Jalabert1990,Marcus1992} localization effects,\cite{Baranger1993,Brezinova2008,Brouwer2008} decoherence\cite{Bird2003,Knezevic2008,Baernthaler2010} and dephasing\cite{Clerk2001,Tanake2002} in ballistic \hyphenation{na-no-struc-tures} nanostructures.
Billiards also serve as ideal systems for the study of the relation of quantum transport to its classical counterpart and the crossover between them.\cite{Baranger1991,Nazmitdinov2002,Richter2002,Fytas2005,Aigner2005}

In all these phenomena the quantum nature of transport manifests itself in interference effects, which arise due to the phase coherence of the underlying scattering process.
The interference of the scattering states and its impact on the inter-lead transmission of a quantum billiard is determined by its geometry including the placement of the leads on its boundary.\cite{Nazmitdinov2002}
For the (semi-) elliptic geometry considered here, the classical dynamics of the closed billiard is regular, with ballistic particle trajectories divided into so called librators and rotators, which intersect its major axis at the segment between the foci and the segments between the foci and the boundary, respectively.\cite{Lenz2007b}
Librators and rotators correspond to quantum eigenmodes\cite{Waalkens1997} localized about the minor semi-axis (also called 'bouncing ball modes') and along the elliptic boundary (also called 'whispering gallery modes'), respectively.\cite{Nazmitdinov2001}
Attaching leads to the straight boundary of the semi-ellipse results in a generalized open mushroom\cite{mushroom} (Bunimovich) billiard with multiple stems of infinite length.
The chaotic trajectories of the closed mushroom, entering its stems, escape into the leads in the open billiard and contribute to transmission.

In the presence of a magnetic vector potential the Aharonov-Bohm (AB) effect\cite{Aharonov1959} induces intricate changes with respect to the interference patterns. Generalized from its occurrence in quantum rings\cite{Frustaglia2001,Kalman2008} to spatially extended billiards,\cite{Wang1994,Rotter2003} it leads to fluctuations of the transmission coefficients with the magnetic field strength.
At high field strengths, where unconfined electrons would occupy localized Landau states, in the billiard edge states form\cite{Beenakker1991} which mediate the propagation along the boundary, with their interference at the lead openings leading to oscillatory transmission behavior.\cite{Rotter2003}

In the present article we investigate the transmission behavior of a 4-terminal semi-elliptic quantum billiard in dependence of its geometrical characteristics and examine how an injected wave can be guided selectively to a chosen output terminal using a magnetic field.
It is shown that the zero-field transmittivity between leads attached to the billiard highly depends on the accessibility, through their coupling to these leads, of librator and rotator modes.
The crucial role of the convex billiard boundary in the transmittivity of such states has been described in Ref.~\onlinecite{Nazmitdinov2001} for a 2-terminal semi-circular billiard.
We in turn calculate the 4-terminal transmission coefficients for varying eccentricity of the semi-ellipse, exploring the dependence of the multilead transmittivity on the curvature of the boundary.
The crossover from librator to rotator modes being the dominant transmission mediators is revealed by altering the lead positions along the straight part of the billiard boundary.
It is shown that an optimal transmittivity between both pairs of symmetrically placed leads can be achieved by an appropriate choice of eccentricity and lead positioning.
For small eccentricity, the placement of attached leads at the foci of the ellipse yields a high corresponding overall transmission coefficient, as would be expected classically.
As a result of interference, though, this condition does not apply for a generic setup, which implies a departure from purely classical considerations.
The importance of the rotator and librator modes is further assessed by gradually perturbing the billiard with a circular disk placed on the curved boundary or in the interior.

With restrictions due to symmetry, the magnetic field changes each transmission coefficient differently; 
the question thus arises whether the multiterminal setup can function as a selective switch between the terminals by tuning the magnetic field, in the sense that transmission of an incoming particle is efficiently favored to specific leads and suppressed to others.
Calculating the multiterminal transmission coefficients of selected setups for varying magnetic field, we show that such output controllability 
is indeed achieved:
it results from the form of the scattering wave function determined by the billiard geometry and the magnetic field.\cite{note_classical}

The paper is organized as follows.
In Section \ref{method} the geometrical setup of the 2D billiard is specified, the theoretical framework for quantum transport is outlined and the computational approach is presented.
Section \ref{symmetries}
summarizes the consequences of the symmetries of the system with respect to the multiterminal transmission.
In Section \ref{transmission_spectra} the main features of the obtained multilead transmission spectra are discussed, along with a description of the underlying mechanisms. 
This is followed by an analysis of the mean transmission components in dependence of the geometric properties of the billiard in Section \ref{geometry}.
The impact of the magnetic field on transport is discussed in Section \ref{magnetic}, concluding on the induced controllability of transmission to selected output leads.
The necessity of the billiard properties for controllable combined output is demonstrated in Section \ref{wires}, and finally its functionality is shown in Section \ref{conductance} in terms of the linear conductance at finite temperature.
Section \ref{summary} contains a brief summary of our conclusions.

\begin{figure}[t!]
    \begin{center}  
      \includegraphics[width=.7\columnwidth]{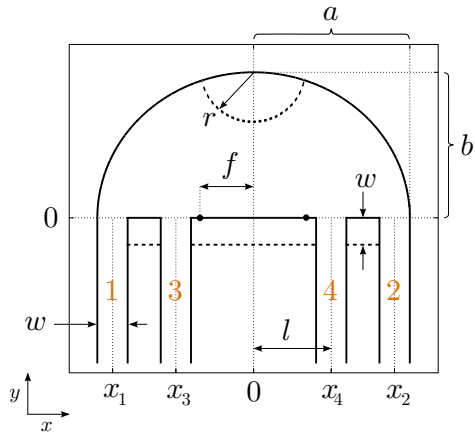}
     \end{center}
  \caption{(Color online) Geometry of the open billiard with indicated length parameters and lead labeling for eccentricity $\epsilon = f / a = \sqrt{1 - b^2 / a^2} = 0.35$, lead width $w = a / 5$, and inner lead positions $\pm l = \pm a / 2$. The two dots at $(x,y) = (\pm f, 0)$ are the foci of the (semi-) ellipse and the dashed lines correspond to alternative setups (see text).}
  \label{sketch}
\end{figure}

\section{Setup and computational approach \label{method}}

The geometry of the 4-terminal hard-wall billiard is shown in Fig.~\ref{sketch}.
It consists of a semi-ellipse of eccentricity $\epsilon = \sqrt{1 - b^2 / a^2}$, where $a$ and $b$ are the major and minor semi-axes respectively, on the straight border of which four vertical semi-infinite leads of equal width $w < a/2$ are attached, symmetrically about the minor axis of the ellipse $x = 0$. 
The elliptic boundary is smoothly continued into the outer leads $1$ and $2$, between which the inner leads $3$ and $4$ are centered at distance $l$ from  the origin.
Semi-infinite leads attached to the closed billiard represent the coupling to particle reservoirs, from which there is no reflection back into the billiard.

To calculate the multi-terminal transmission coefficients of the system, Dirichlet boundary conditions are imposed on the scattering wave function along the boundary of the billiard, defined by a hard-wall potential $V(\bf{r})$.\cite{softwall}
Adapting to natural units, we set $\hbar = - e = m = 1$, where $e < 0$ and $m$ denote the charge and effective mass, respectively, of a spinless particle.
The single-particle Hamiltonian is then written
\begin{equation}
H = \frac{1}{2} \left(\,{\bf{k}} + {\bf{A}} \right)^2 + V ~~,
\end{equation}
where the vector potential $\bf{A}$ generates a magnetic field ${\bf{B}} = \nabla \times {\bf{A}} = B \hat{\bf{z}}$ perpendicular to the plane of the structure,
with strength $B$ that is homogeneous over the extent of the billiard and falls off linearly to zero over a length $\approx 30 w$  in the attached leads.
The particle is incident in one of the four leads with energy
\begin{equation}
E = \frac{{\bf{k}}^2}{2} = \frac{1}{2} \left[ (k^n_y)^2 + \left(\frac{n \pi}{w}\right)^2 \right] = \frac{1}{2} \left(\frac{\pi}{w} \right)^2 \kappa^2~~,
\label{energy}
\end{equation}
where $n = 1, 2, ...$ labels the subbands of the longitudinal momentum $k^n_y$ along the unperturbed leads, generated by the transversal confinement to their common width $w$.
The scaled momentum $\kappa  = k \cdot w/\pi = \sqrt{2E } \cdot w/\pi$ thus varies continuously in the interval $n < \kappa < n+1$ for motion in the $n$-th subband.

The Hamiltonian is discretized on a square grid of unit lattice constant in the tight-binding approach, with the magnetic vector potential incorporated through the Peierls substitution.\cite{Peierls}
Considering (spin polarized) electronic transport at a GaAs/AlGaAs ($m = 0.069~ m_e$) interface and setting the lattice constant unit to $\alpha = 2\rm~{nm}$, the unit of energy becomes $\hbar / m \alpha^2 = 0.276~\rm{meV}$  and the unit of field strength $\hbar / |e|\alpha^2 = 164.55~\rm{T}$, while the lengths in the system are scaled by a reference ellipse major semi-axis $a = 100\alpha = 200~\rm{nm}$.

The coupling of the system to the external semi-infinite leads $i = 1, 2, 3, 4$ attached to the system is described by self-energies ${\bf \Sigma}_i$, which are analytically obtained for $B = 0$ and contribute non-Hermitian blocks to the Hamiltonian matrix.\cite{Datta1995}
From the single-particle (retarded) Green function of the system
\begin{equation}
{\bf G}(E) =[ E{\bf I} - ( {\bf H} + {\bf \Sigma}) ]^{-1} ~~
\end{equation}
where
\begin{equation}
{\bf \Sigma}(E) = \displaystyle \sum_i {\bf \Sigma}_i ~~,
\end{equation}
the part ${\bf G}_{ij}$ describing the propagation from lead $j$ to lead $i$ 
is computed using a parallel implementation of the recursive Green function method (RGM),
where a decomposition scheme among communicating processors allows for the computation to be done in a parallel manner.\cite{Drouvelis2006}

Since in general $B \neq 0$, the self-energies are addressed to grid points in the leads where the field strength has decreased to zero.
The computation of ${\bf G}$ is then more efficient if the system is decomposed into parts, so called modules, of increased symmetry,
by utilizing our own implementation of a modular version of the RGM, introduced in Ref.~\onlinecite{Rotter2000}.
The considered setup is thus assembled by two types of modules:
the semi-elliptic scatterer at constant magnetic field strength $B$, and a lead part with the magnetic field decreasing linearly to zero.
The modules are then subsequently connected to form the complete scatterer, and the Green function of the connected modules is obtained in each step by solving a matrix Dyson equation in tight-binding form.\cite{Rotter2000}

The scattering matrix elements connecting the leads of the system are expressed in terms of the Green function by the Fisher-Lee relation,\cite{Fisher1981}
which leads to a compact trace formula for the multiterminal transmission coefficients,\cite{Datta1995}
\begin{equation}
T_{ij}(E) = Tr \left({\bf  \Gamma}_i {\bf G} {\bf \Gamma}_j {\bf G}^\dagger \right) ~~ (i \neq j) ~~,
\end{equation}
with ${\bf \Gamma}_j = i ( {\bf \Sigma}_j - {\bf \Sigma}^\dagger_j)$.
For $i = j$, the reflection coefficient of each lead $j$, $T_{jj} \equiv R_j$, is given by the sum rule
\begin{eqnarray}
n(E) = \sum_i{T_{ji}(E)} & = & \sum_i{T_{ij}(E)} \nonumber \\ 
		\cr	 & = & \sum_{i \neq j}{T_{ij}(E)} + R_j(E) ~~,
\label{sum_rule}
\end{eqnarray}
where $n(E) = int[\kappa(E)]$ is the number of open channels in the leads (of common width $w$) at energy $E$, $int[~~]$ denoting the integer part.
This sum rule results, for coherent transport, from the unitarity of the scattering matrix, which in turn is a consequence of probability flux conservation.

The Green function of the system is also used to calculate the local density of states (LDOS) at site $\bf{r}$ through the relation
\begin{equation}
\rho({\bf{r}},E) = \frac{1}{2\pi} \langle{\bf{r}}\vert{\bf{\Phi}}(E)\vert{\bf{r}}\rangle ~~,
\end{equation}
where $\bf{\Phi}={\bf{G}}{\bf{\Gamma}}{\bf{G}}^\dagger$ 
is the spectral function and ${\bf{\Gamma}}$ generally a weighted sum over the ${\bf \Gamma}_{i}$
according to the distributions of incoming states in the leads (the Fermi distribution in the case of electrons). 
In the cases presented here, we have chosen ${\bf{\Gamma}} = {\bf \Gamma}_{i}$ for a certain lead $i$,
so that $\rho({\bf r},E)$ corresponds to the probability density resulting from an incoming monochromatic wave of energy $E$ in lead $i$.

\section{Results and discussion\label{results}}

\subsection{Symmetries of the transmission coefficients\label{symmetries}}

Before investigating the multiterminal transmission coefficients in varying geometry and field,
we show how symmetries present in the system can be used to reduce the number of independent coefficients to be calculated.
Time reversal symmetry (TRS) yields a Hermitian conjugated scattering matrix under inversion of the magnetic field, implying for the transmission coefficients:\cite{Datta1995}
\begin{equation}
T_{ij}(E;B) = T_{ji}(E;-B)
\label{TRS}
\end{equation}
This reciprocity relation halves the number of independent transmission coefficients $T_{i \neq j,j}$ when both field directions are considered.
The reflection coefficients $R_j = T_{i = j,j}$ are given by the sum rule Eq.~(\ref{sum_rule}) and remain the same under field reversal [cf. Eq.~(\ref{TRS})], reducing the number of independent coefficients by the number of leads.
For a 4-terminal billiard the $4 \times 4 = 16$ coefficients are thus reduced to 6 independent ones.

In our billiard the spatial reflection symmetry about the $y$-axis introduces additional relations between symmetric pairs of leads.
If leads $i$ (at $x_i$) and $j$ (at $x_j$) are placed symmetrically to leads $i'$ (at $-x_i$) and $j'$ (at $-x_j$), then
\begin{equation}
T_{ij}(E;B) = T_{i'j'}(E;-B) ~~,
\label{reflection}
\end{equation}
as the equations of motion for a (charged) particle are invariant under the transformation $(x,B) \rightarrow (-x,-B)$ in the symmetric billiard.
Explicitly, we get the two additional relations $T_{24}(B) = T_{13}(-B)$ and $T_{41}(B) = T_{32}(-B)$, reducing the number of independent coefficients to 4 (if $i' = j$ and $j' = i$, this reflection symmetry coincides with the TRS).

\begin{table}[t]
    \begin{center}
     \hspace{-.5cm}
      \includegraphics[width=.9\columnwidth]{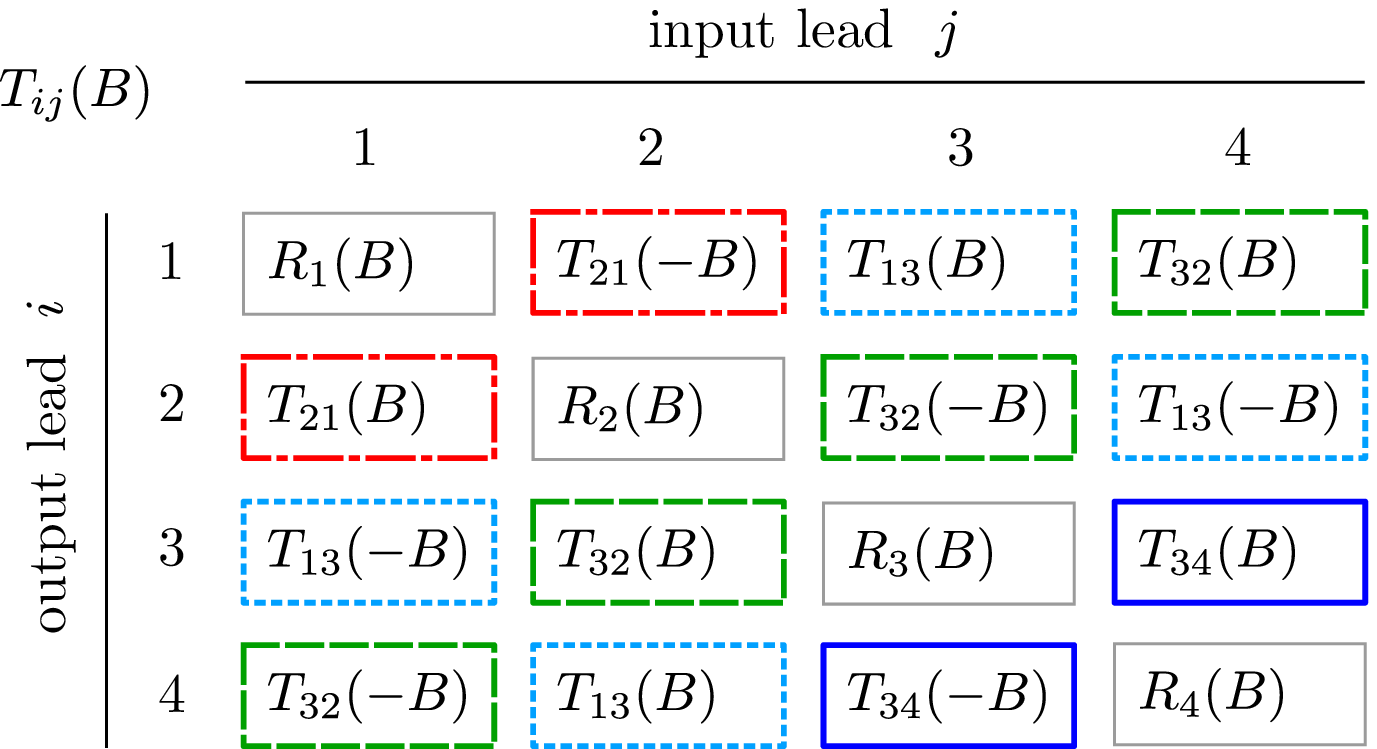}
     \end{center}
\caption{Multiterminal transmission coefficients $T_{ij}(B)$ from lead $j$ (columns) to lead $i$ (rows), deduced from the coefficients $T_{21}$, $T_{32}$, $T_{13}$ and $T_{34}$ at magnetic field strength $B$ (see text). The surrounding boxes of the $T_{i \neq j, j}$ correspond to the plotted line types in Figs.~\ref{Tmean_rvar} -- \ref{Tmean_B_wires}. 
} 
\label{symmetry_table}
\end{table}

In the following, we will work explicitly with the coefficients  $T_{21}$, $T_{32}$, $T_{13}$ and $T_{34}$ with input in each one of the four leads, because this set serves best for our discussion of the results.
In Table~\ref{symmetry_table} all transmission coefficients $T_{ij}(B)$ are explicitly expressed in terms of these four.

\subsection{Transmission spectra at zero magnetic field\label{transmission_spectra}}

\begin{figure}[t!]
    \begin{center}  
      \includegraphics[width=\columnwidth]{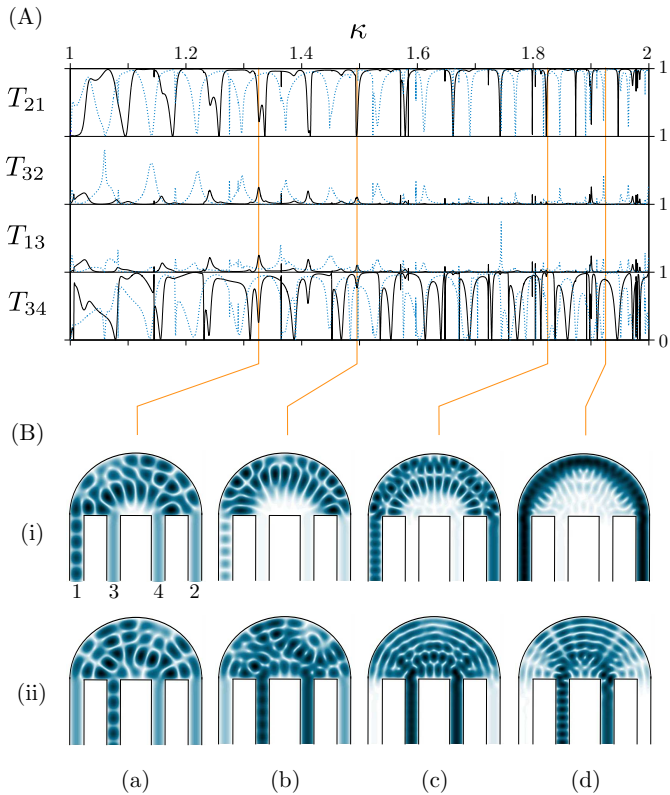}
     \end{center}
      \caption{(Color online) (A) The four independent transmission coefficients $T_{ij}$ at zero magnetic field, as a function of the scaled energy $\kappa = k \cdot w/\pi = \sqrt{2E} \cdot w/\pi$, for the elliptic ($\epsilon = \epsilon_3 = 0.35$, black solid line) and circular (blue dotted line) billiard border, both with $w = a / 5$ and $l = \epsilon_3 a$. (B) LDOS for the semi-elliptic billiard at different energies (a), (b), (c), (d) indicated by the vertical lines in (A), with the particle incident in the (i) outer and (ii) inner left lead. The colormap scales with $\sqrt{\rho(x,y)}$ from white ($\rho = 0$) to black ($\rho = \text{max}$) and is normalized to its maximal value in each plot.}
  \label{Tspec_LDOS}
\end{figure}

The generic features of the zero-field multiterminal transmission spectra are presented for a geometric setup with relatively high overall transmission between inner leads, since this will prove to be a key property for the desired controllability of output terminal.
The $T_{ij}(\kappa)$ are shown in Fig.~\ref{Tspec_LDOS}(A) for $\kappa$ within the first channel ($n = 1$).
All terminal combinations are represented, since their relations in Table~\ref{symmetry_table} simplify accordingly for $B = -B$.
Fig.~\ref{Tspec_LDOS}(B) shows the LDOS at selected energies for different leads of incident wave.
Note that the presence of interference fringes in a lead signifies back-reflection into that lead.

As we see in $T_{21}(\kappa)$, transmission is overall close to unity between the outer leads.
This results from rotator modes of the semi-elliptic billiard that would leak into finite stems at the outer lead positions, and therefore are strongly coupled to these leads in the open system.
These leaking states constitute non-resonant pathways for transport, whose superposition leads to a high transmission background in $T_{21}$ [see LDOS in Fig.~\ref{Tspec_LDOS}(B)(i,d)], smoothly varying in energy.\cite{Buchholz2008,Morfonios2009}

States weakly coupled to the outer leads [see LDOS in Fig.~\ref{Tspec_LDOS}(B)~(i,a) and (i,b)] constitute resonant pathways, whose interference with the non-resonant pathways leads to sharp resonances in $T_{21}(\kappa)$ of width proportional to their coupling, which possess the characteristic Fano lineshape asymmetry.\cite{Fano1961,Nockel1994,Gores2000}
In the case of a single non-resonant pathway, the asymmetry is caused by a transmission zero close to the resonant energy, owing to complete destructive interference between the resonant and non-resonant state.
In our system each quasi-bound state (resonant pathway) in general couples to multiple multiterminal leaking states (non-resonant pathways), which renders the total interference partially destructive and thus raises the minimum of the Fano resonance from zero.\cite{Fano1961}
Corresponding to the eigenstates of the closed semi-ellipse, the resonances superimposed on the transmission background appear in series of different quasi-periodicity in $\kappa$, determined by the quantization of the wave-number of the semi-elliptic modes (in analogy with the detailed description in Ref.~\onlinecite{Buchholz2008} for the oval billiard).

The coefficient $T_{34}$ is also overall high in the first channel for the chosen eccentricity and lead positioning, in this case resulting from the strong coupling mostly of librator modes to the inner leads. 
The convexity of the boundary plays a crucial role for this behavior, since it focuses the scattering wave function, incident in an inner lead, around the middle of the straight boundary of the billiard [see Fig.~\ref{Tspec_LDOS}(B)(ii,c) at which energy $T_{34}$ practically reaches unity].
Note that, since the leads in this setup are centered at the foci of the ellipse, classically both librators and rotators intersect the lead openings; quantum mechanically though, there is a larger number of eigenmodes of librator type with maxima at the foci.\cite{Waalkens1997}
Due to interference between these modes, setting $l = f = \epsilon a$ is not a necessity for the acquired high inner-lead transmission; it depends also on the chosen $\epsilon$, as will be shown in Section \ref{geometry}.

In $T_{34}$ the sharp resonant dips in the high background are at the same positions as in $T_{21}$ but generally of different width, arising from the same quasi-bound states coupling with different strength to the inner leads.
Also, the Fano minima in $T_{34}$ are of different height than in $T_{21}$, since different non-resonant transport paths are provided by the librator modes, interfering with the resonant states.
For some sharp resonances in $T_{34}$ (e.g. at $\kappa \approx 1.145$ and $1.365$) the Fano asymmetry is more distinct, as they lie within a dip of the transmission background.

As a consequence of probability flux conservation [Eq.~(\ref{sum_rule})], unit transmission in either $T_{21} = T_{12}$ or $T_{34} = T_{43}$ leads to vanishing transmission in both $T_{32} = T_{23} = T_{41} = T_{14}$ and $T_{13} = T_{31} = T_{24} = T_{42}$.
Thus, transmission between an outer and an inner lead, represented by the coefficients $T_{32}$ and $T_{13}$ in Fig.~\ref{Tspec_LDOS}(A), is almost zero over the whole channel for the chosen geometric parameters.
It exhibits resonant peaks, coinciding with dips for the symmetric lead pairs, albeit of rather low amplitude, since the the incoming wave is mostly reflected into the same lead [see Fig.~\ref{Tspec_LDOS}(B)(a)] or transmitted to the symmetrically placed lead [see Fig.~\ref{Tspec_LDOS}(B)(b)].
As each eigenstate of the semi-elliptic billiard is symmetric, it generically couples with different strength to inner and outer leads;
desymmetrized lead positions then lead to lower transmittivity of the corresponding resonant states in the open system.\cite{Nazmitdinov2001}

When the eccentricity of the billiard is slightly changed, the eigenstate wavelength in the elliptic coordinates is accordingly modified\cite{Waalkens1997} and consequently the corresponding transmission resonances shift in $\kappa$.
This is evident in Fig.~\ref{Tspec_LDOS}(A) for the semi-circular billiard (dotted cyan line): its area is slightly larger than the semi-ellipse, so that the resonances are shifted to lower $\kappa$.
The dips in $T_{21}$ and $T_{34}$ (or peaks in $T_{32}$ and $T_{13}$) are overall broader for the semi-circular billiard, owing to the enhanced coupling of its eigenstates to both inner and outer leads.
It also shows a pronounced imbalance between $T_{32}$ and $T_{13}$ at low energy:
resonant transmission is larger from an outer lead to the inner lead which is closer to the opposite outer lead, where rotator modes can be accessed.

\subsection{Geometry-dependent mean transmission\label{geometry}}

Having presented the zero-field spectral features and their origins, we proceed to investigate the overall multiterminal transmittivity in dependence of the geometry of the setup by computing, for each set of parameters, the channel-integrated mean transmission coefficients
\begin{equation}
\bar{T}^{(n)}_{ij} = \int_{\kappa = n}^{\kappa = n + 1} T_{ij}(\kappa) ~~
\label{mean_transmission}
\end{equation}
of the $n$-th transversal subband, which constitute a measure of the overall response of the system upon an incoming wave in one of the leads.

\begin{figure}[t!]
    \begin{center}  
      \includegraphics[width=.8\columnwidth]{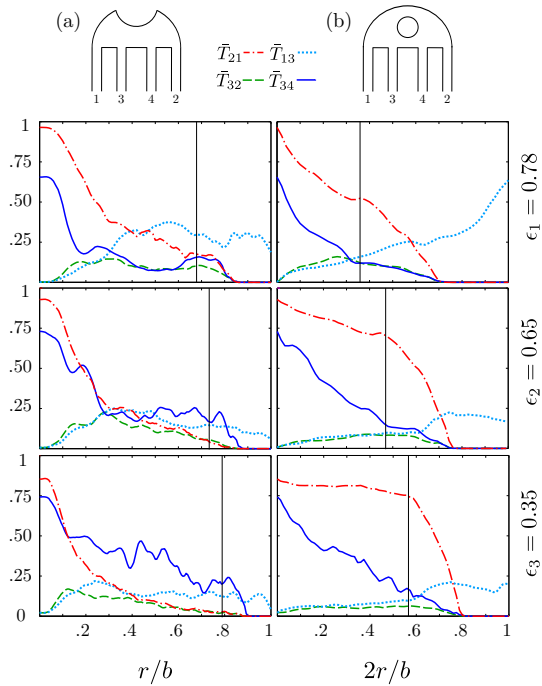}
     \end{center}
      \caption{(Color online) Zero-field channel-averaged transmission coefficients $\bar T_{ij}$ between terminals $(i,j) = (2,1)$ (red dashed-dotted line), $(3,4)$ (blue solid line), $(1,3)$ (cyan dotted line), $(3,2)$ (green dashed line) with lead widths and positions as in Fig.~\ref{Tspec_LDOS} and eccentricities (from top to bottom row) $\epsilon_1 = 0.78$, $\epsilon_2 = 0.65$, $\epsilon_3 = 0.35$, for the semi-elliptic billiard with a disk inserted in the billiard (a) centered at $(x, y) = (0, b)$ with radius $r$ varying from $0$ to $b$ and (b) centered at $(x, y) = (0, b/2)$ with $r$ varying from $0$ to $b/2$. The vertical lines denote the threshold radius $r_t$ for tunneling within the first channel (see text).}
  \label{Tmean_rvar}
\end{figure}

In order to confirm the role of rotator and librator modes in coupling symmetrically placed leads we first explore the effect of a perturbing disk on the mean transmission coefficients.
In Fig.~\ref{Tmean_rvar} the $\bar{T}^{(n = 1)}_{ij} \equiv \bar{T}_{ij}$ (the channel superscript is dropped for $n = 1$) are shown as a function of varying radius of the disk, for three different eccentricities.
In Fig.~\ref{Tmean_rvar}~(a) the disk constitutes a circular recess of the elliptic boundary, and in Fig.~\ref{Tmean_rvar}~(b) it leaves the boundary of the billiard unperturbed, but partially blocks direct transport in its bulk.

In case (a) the rotator modes are gradually destroyed with increasing $r$, because incoming waves from an outer lead are deflected on the concave part of the boundary into the billiard interior (an analogous situation is presented in Ref.~\onlinecite{Nazmitdinov2001} for a fixed rectangular cut).
Similarly, the librator modes are destroyed since they rely on the focusing ability, predominantly around the $y$-axis, of the convex boundary.
As a result, the direct pathways between symmetric leads are depleted, leading to an abrupt decrease in $\bar{T}_{21}(r)$ and $\bar{T}_{34}(r)$ above a critical disk radius, which is about $r_c \approx w/3$ for all $\epsilon$ [note that the scaling $r/b$ stretches the plots horizontally for larger $\epsilon$: $r_c (\epsilon_1)\approx 0.10\,b$, $r_c (\epsilon_2)\approx 0.08\,b, r_c (\epsilon_3)\approx 0.06\,b$].
Below this critical disk radius, efficient guiding of rotator modes and focusing of librator modes can be considered robust to boundary perturbations.\cite{perturb}
For smaller $\epsilon$, $\bar{T}_{34}(r)$ remains substantial over larger $r$, since the larger billiard can support a larger number of direct scattering pathways between the inner leads.
Deflection on the perturbed boundary enhances scattering to asymmetrically placed leads, leading to increased $\bar{T}_{32}$ and $\bar{T}_{13}$ from zero for $r \neq 0$.

In case (b) the librator modes are again rapidly destroyed by the disk, resulting in a decrease of $\bar{T}_{34}(r)$ similar to that in (a).
Rotator modes sufficiently localized along the elliptic boundary survive up to some disk size and still connect the outer leads;
especially in (b, $\epsilon_3$), $\bar{T}_{21}(r)$ forms a characteristic high plateau until decreasing abruptly when the remaining free width between disk and boundary becomes smaller than the leadwidth $w = a/5$.

The latter condition is met above a threshold radius $r_t$ given by (a) $r_t/b = 1 - (5\sqrt{1 - \epsilon^2})^{-1}$ and (b) $2r_t/b = 1 - 2(5\sqrt{1 - \epsilon^2})^{-1}$, denoted by vertical lines in Fig.~\ref{Tmean_rvar}, at which the energy threshold for transport between opposite sides of the disk enters the first channel in the leads.
When this threshold rises above $\kappa = 2$ (for free width smaller than $w/2$), only tunneling through the constrictions contributes to $\bar{T}_{21}$, $\bar{T}_{34}$ and $\bar{T}_{32}$, which then practically vanish; 
in contrast, $\bar{T}_{13}$ is enhanced for large $r$, particularly in case (b, $\epsilon_1$).
Furthermore, larger number of accessible eigenstates causes more resonant features in the transmission spectra, leading to increased fluctuations of the $\bar{T}_{ij}$ in continuously varying geometry for larger billiards (row $\epsilon_3$).

\begin{figure}[t!]
    \begin{center}  
      \includegraphics[width=\columnwidth]{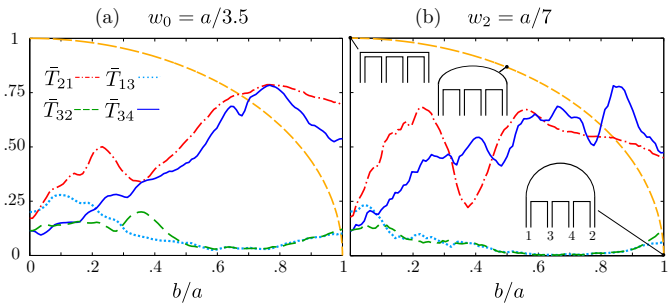}
     \end{center}
      \caption{(Color online) $\bar T_{ij}$ for varying ratio $b/a$, with equidistantly attached leads of  width (a) $w_0 = a / 3.5$ and (b) $w_2 = a / 7$. The light (orange) dashed line shows the eccentricity $\epsilon$. The lower horizontal edge of the billiard here lies at $y = -w$ (see Fig.~\ref{sketch}), and the upper edge is varied from straight ($b = 0$) to elliptic ($0 < b < a$) to circular ($b = a$), as illustrated by the inset pictures in (b).}
  \label{Tmean_bvar}
\end{figure}

Librator- and rotator-like eigenmodes of the billiard were shown to be necessary for high inner and outer lead transmission;
nevertheless, their coupling to the leads further depends on the eccentricity of the unperturbed billiard for a given lead positioning.
In Fig.~\ref{Tmean_bvar} the $\bar{T}_{ij}$ are plotted against the ratio $b/a = \sqrt{1-\epsilon^2}$ for different $w$.
In order to access the limit of zero curvature ($\epsilon = 1$), the straight edge of the billiard is lowered by one leadwidth (see dashed line in Fig.~\ref{sketch}).
For the true semi-elliptic setup the features in $\bar{T}_{ij}(b/a)$ are shifted to larger $b$ (and thus smaller $\epsilon$), so that the change in size is compensated and the corresponding eigenmodes remain approximately at the same energies.
$\bar{T}_{21}$ and $\bar{T}_{34}$ overall increase with $b$, as rotator- and librator-like modes start to form which couple outer and inner leads;
in contrast, $\bar{T}_{32}$ and $\bar{T}_{13}$ overall decrease and possess a common broad minimum.
For a certain eccentricity range a separation between inner and outer leads is thus possible, in the sense that cross-coupling between them (i.e. between an outer and an inner lead) is almost eliminated in zero magnetic field.
Further, depending on $\epsilon$, rotator-like modes leaking into the outer leads can interfere into a suppressed transmission background, causing the characteristic wide dip in $\bar{T}_{21}$ for $b \approx a/3$.
By decreasing $w$ the transversal subbands are shifted up in energy and the wavelength of the incoming particle decreases relatively to the billiard size, so that a larger number of eigenstates is spanned within the first channel.
This leads to increased fluctuations of the $\bar{T}_{ij}$ in Fig.~\ref{Tmean_bvar}(b), similar to those in Fig.~\ref{Tmean_rvar} (row $\epsilon_3$).

Let us now investigate the overall transmission behavior in dependence of the placement of the inner leads, which determines their coupling to the different billiard eigenmodes.
As the inner leads are moved away from the center, there is a gradual crossover of the direct transport paths from librator- to rotator-type states.
The question arises whether it is sufficient, or even necessary, to place the inner leads at the ellipse foci in order to achieve high transmission between them, as shown in Fig.~\ref{Tspec_LDOS}.
In the classical picture the separation of librators and rotators by the focal points is sharp, and in the limit of zero leadwidth, all trajectories coming in from one focus are scattered directly (by only one reflection at the elliptic boundary) to the other, leading to unit transmission.
For a finite leadwidth, a portion of the incoming trajectories is scattered onto the straight segments between the leads and eventually into an outer lead, so that the inner lead transmission is lowered from unity.
In the quantum case, additionally, the spatial separation between librator and rotator modes is not sharp, especially at the low energies considered;
thus, however narrow, the inner leads couple to both types of modes.
Most importantly, though, the transmission coefficients highly depend on interference phenomena between the resonant states coupling to the leads.
Even if the inner leads are placed close to the foci, where most eigenmodes possess a probability maximum,\cite{Waalkens1997} multiple destructive interference between them may lead to low overall transmission.

\begin{figure}[t!]
    \begin{center}  
      \includegraphics[width=\columnwidth]{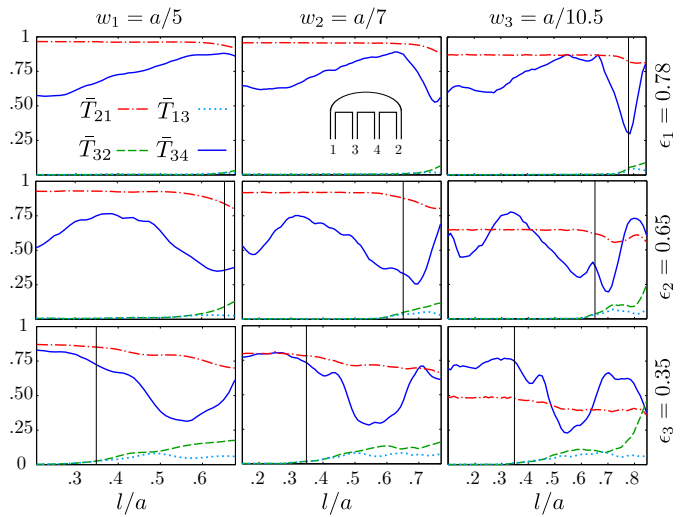}
     \end{center}
      \caption{(Color online) $\bar T_{ij}$ in the unperturbed semi-elliptic billiard, for lead width (left to right column) $w_1 = a / 5$, $w_2 = a / 7$, $w_3 = a / 10.5$ and eccentricities (top to bottom row) $\epsilon_1 = 0.78$, $\epsilon_2 = 0.65$, $\epsilon_3 = 0.35$, as a function of the displacement $l$ of the inner leads from the origin. The vertical lines show the position of the focal points $f/a = \epsilon$ for each setup.}
  \label{Tmean_lvar}
\end{figure}

In Fig.~\ref{Tmean_lvar} the variation of the mean transmission components with the inner lead displacement $l$ is shown for different leadwidths and eccentricities.
For large $\epsilon$, $\bar{T}_{34}(l)$ increases to maximum when the inner leads are next to the outer ones, though with transport dominated by librator modes, since the foci lie within the outer lead openings ($w_1$, $\epsilon_1$).
This trend is inverted for smaller $\epsilon$, where the foci come closer to the origin and allow for the coupling to rotator modes.
Then ($w_1$, $\epsilon_3$) $\bar{T}_{34}(l)$ is maximal for the inner leads close to the origin and decreases to minimum for large $l$, as a result of destructive interference.
For narrower leads (columns $w_2$ and $w_3$) these features remain, with enhanced fluctuations like in Fig.~\ref{Tmean_bvar}.

We indeed see that placing the inner leads close to the foci does not necessarily lead to high overall transmission between these leads, demonstrating the departure from the classically expected behavior of our billiard in the deep quantum regime.
Some of the setups, in particular of Fig.~\ref{Tmean_lvar}~($w_3$, $\epsilon_1$), even exhibit a wide minimum in $\bar{T}_{34}$ around $l = f$.

$\bar{T}_{21}(l)$ overall increases with $w$ and $\epsilon$, but remains largely unaffected by the variations in $l$.
It decreases slightly only when $l$ is large enough for the inner leads to couple to the same modes as the outer leads, which causes a corresponding increase in $\bar{T}_{32}$ and $\bar{T}_{13}$.

For the magnetic control of multiterminal transmission, to be discussed in the following section, it is important to achieve high inner- and outer-lead zero-field transmission, while cross-coupling is suppressed;
we see that these conditions are met in our setup by combining small $l$ and $\epsilon$ with relatively large $w$.

\subsection{Transmission in a magnetic field \label{magnetic}}

To understand how conductivity between terminals can be selectively manipulated with the magnetic field, we first analyze its impact on the transmission spectra and its interplay with the geometric properties. 
In the presence of the field the resonant states accordingly shift in energy,\cite{Nockel1992} while its influence on their phase modifies their coupling to the leads and the interference with other states.
Therefore, the widths of sharp Fano resonances generally change, and the non-resonant pathways interfere into a different transmission background.
If the field is very weak, the spatial distribution of the eigenmodes remains practically unaffected, as does their individual coupling to the lead openings.
A drastic change in the overall transmission in a weak field can still take place, though, when a small number of leaking modes interfere.\cite{Buchholz2008}
For a stronger magnetic field the spatial distribution of the states changes enough to generally yield a completely modified transmission spectrum.
In the classical picture the charged particle moving in the billiard is deflected into circular orbit of gyromagnetic (Larmor) radius $r_L = k/|B|$, making the classification of trajectories into rotators and librators inapplicable.
When the field strength is further increased, $r_L$ eventually becomes so small that the particle moves along skipping trajectories at the billiard edges.\cite{Beenakker1991}
The corresponding quantum scattering wave function is localized into edge states,\cite{Rotter2003} which enable almost reflectionless transport along the boundary.

\begin{figure}[t!]
    \begin{center}  
      \includegraphics[width=1\columnwidth]{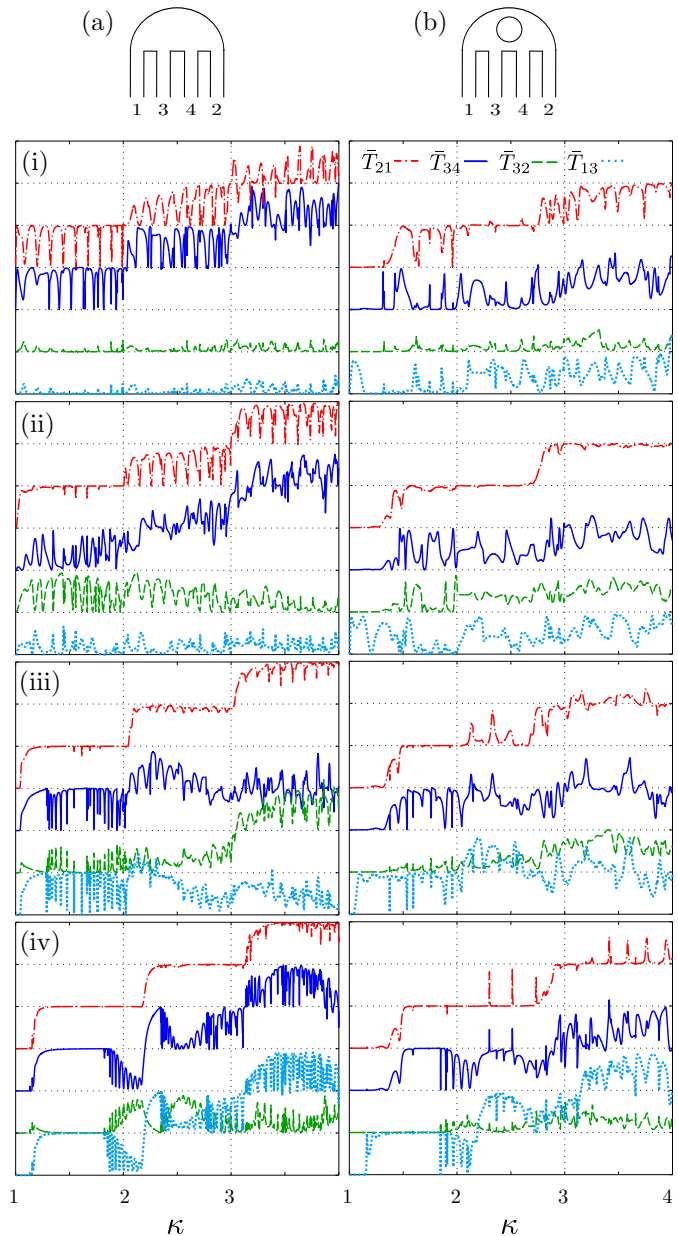}
     \end{center}
      \caption{(Color online) Transmission coefficients $T_{21}$ (red dashed-dotted line), $T_{34}$ (blue solid line), $T_{32}$ (green dashed line), $T_{13}$ (cyan dotted line), with offsets $3, 2, 1, 0$ respectively, as a function of $\kappa$ within the first three transversal subbands $n = 1, 2, 3$ of leads of width $w = w_0 = a / 3.5$, attached equidistantly with $l = (a - w/2)/3 = 0.84\,f$, (a) for the unperturbed semi-ellipse billiard with $\epsilon = 0.35$ and (b) for the same geometry with a disk of radius $r = b/2 - 2w/3$ centered at $(0, b/2)$, at magnetic field (i) $B = 0$, (ii) $B = 0.002$, (iii) $B = 0.005$, (iv) $B = 0.010$, with ${\bf{B}} = B \hat{\bf{z}}$ pointing outwards from the billiard plane.\\}
  \label{Tspec_B}
\end{figure}

\begin{SCfigure*}
    \centering 
      \hspace{-0.3cm}
      \includegraphics[width=0.79\textwidth]{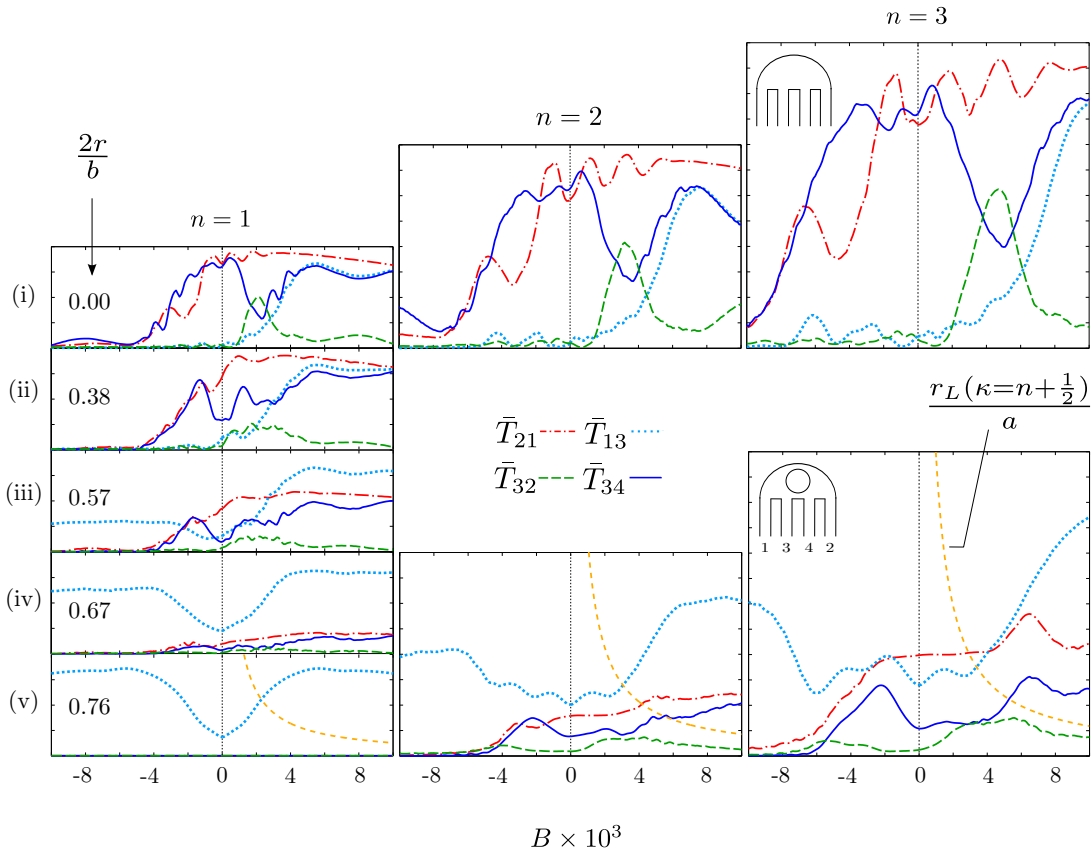}
      \hspace{.3cm}
      \caption{(Color online) Mean transmission coefficients $\bar T_{ij}$ as a function of the applied magnetic field $B$ for the billiard of Fig.~\ref{Tspec_B} with increasing disk radius $r$ in the $n = 1$ subband (top to bottom panels in left column); for zero and maximal $r$ (top and bottom row, respectively) also the $n = 2$ and $n = 3$ subbands are shown. Each plot is of height $max[\bar T_{ij}^{(n)}] = n$. The light (orange) dashed line in the bottom row panels shows the scaled Larmor radius $r_L/a$ at the $n$-th channel's center $\kappa = n + 1/2$. \vspace{1.8cm}}
  \label{Tmean_Bvar}
\end{SCfigure*}

In Fig.~\ref{Tspec_B}(a) the transmission spectra at (i) zero, (ii - iii) intermediate and (iv) high magnetic field strength are shown for the unperturbed billiard.
As in Sec.~\ref{transmission_spectra}, the geometry is adjusted for high inner- and outer-lead zero-field transmission;
in order to concentrate on magnetically induced spectral changes, even wider leads are used, avoiding increased fluctuations from multiple interference of billiard eigenmodes.
The three first channels $n = 1, 2, 3$ are addressed, showing the typical stepwise increase of (maximal) transmission with $n$ and the effect of channel mode coupling.
$T_{21}$ and $T_{34}$ are overall high in the first channel at $B = 0$ [(a)(iii)], as previously discussed, but get modulated in the higher channels by increasingly wide dips and lowered background transmission caused by multimode interference.
The field strength for (a)(ii) is chosen to suppress inner-lead transmission, which enables the control of multiterminal transport, as will be shown in Sec.~\ref{conductance}.
Waves coming in from lead $1$ are deflected onto the curved boundary, which decouples them further from the inner leads, reducing the width of the dips in $T_{21}$.
$T_{34}$ is overall lowered, since waves coming in from lead $4$ (containing the right ellipse focus) are no longer focused into lead $3$: the librator modes are destroyed in the presence of the deflecting field.
On the contrary, $T_{32} = T_{14}$ is drastically increased and $T_{12}$ (not shown) is accordingly reduced: the Larmor radius at this field is, approximately, the one needed to deflect classical trajectories from lead $4$ into lead $1$ (or from lead $2$ to lead $3$) without reflection at the boundary, for energies in the first channel.
This condition is not fulfilled anymore at the even higher field strength in (a)(iii), where edge states start to form.
In the first channel, transmission is then favored to the next neighboring lead ($T_{21}$, $T_{34}$ and $T_{13} = T_{42}$) and suppressed between other lead pairs.
The complementarity between the multi-terminal coefficients is here clearly manifest in the coincidence of the dips in $T_{34}$ and $T_{13} = T_{42}$ with the peaks in $T_{32} = T_{14}$.
These resonances appear when the nodal pattern of interfering edge states (or a multiple of the diameter of the classical skipping orbits) matches the distance of leads $2$ and $3$ instead of adjacent leads (this behavior is described in detail in Ref.~\onlinecite{Rotter2003} for a 2-terminal billiard).
In the second and third channel interference of a larger number of accessible modes enhances again fluctuations in the transmission background.
At very high field strength [(a)(iv)] the edge states lie so far apart in energy (and classically $r_L$ is so small) that plateaus of unit transmission to clockwise subsequent leads appear for energies in the lowest magnetoelectric (Landau) subband.
When more edge states are energetically accessible we observe a difference between scattering at smooth and sharp lead openings.
Diffraction at the sharp edges causes mixing and interference of the different edge states,\cite{Rotter2003} leading to oscillations in $T_{34}$, $T_{32}$ and $T_{13}$.
Only the $T_{21}$ coefficient exhibits perfect transmission even at higher energy, since the edge states adiabatically follow the smooth elliptic boundary from lead $1$ to $2$.
The stepwise increase of transmission with $n$, most pronounced in $T_{21}$, is shifted to higher $\kappa$ with increasing field strength, following the threshold energies of the magnetic subbands.

As in Sec.~\ref{geometry}, a disk inside the semi-elliptic billiard drastically changes its transmission properties by blocking direct transport paths between the leads.
In Fig.~\ref{Tspec_B}(b) the field dependent spectra are shown for this setup, with the disk leaving constrictions of minimal width $2w/3$ with the boundary.
The transmission threshold is then essentially shifted from $n$ to $3n/2$, as can be seen by replacing $w$ with $2w/3$ in Eq.~(\ref{energy}).
Thus, $T_{21}$, $T_{34}$, or $T_{32} > n - 1$ below $\kappa = 3n/2$ results from tunneling of the wave function through the constrictions.
Distinct resonant tunneling peaks are seen below this threshold for $n = 2$ (that is, $1.5 < \kappa < 3$) in $T_{21}$ and $T_{34}$ at high field strength [Fig.~\ref{Tspec_B}(b)(iii) and (iv)], mediated by edge states that are localized on the disk edges (similar to the states leading to sharp reflection resonances, due to different geometrical setup, in Ref.~\onlinecite{Szafran2010}).
The effective resonator length is, approximately, the mean periphery $2\pi(r + w/3)$, leading to the observed peak spacing $\Delta\kappa \approx 0.247$ in (b)(iv).
On the other hand, scattering upon the disk favors transmission between leads on the same side of it, so that $T_{13}$ overall increases and qualitatively approaches the unperturbed case for strong fields.

In Fig.~\ref{Tmean_Bvar} the four independent channel mean components $\bar{T}_{ij}$ are plotted as a function of $B$ for different radii $r$ of the inserted disk [rows (i) to (v)].
The $\bar{T}^{(2)}_{ij}$ and $\bar{T}^{(3)}_{ij}$ are shown for $r = 0$ (unperturbed billiard) and $r = b/2 - w/3$ (almost divided billiard), where also the Larmor radius $r_L(B) = \pi/w \cdot (n + \frac{1}{2})/|B|$ at each channel center is plotted to show the field impact on the classical trajectories.
In absence of the disk [row (i)], $\bar{T}_{21}$ is close to unity for $B \geqslant 0$, falling off slowly at large $B$ as the magnetic threshold enters the channel [see Fig.~\ref{Tspec_B}(a)(iv)].
For $B < 0$ it decreases abruptly with field strength when $r_L < a$, since the incoming waves in lead $1$ are deflected away from the elliptic boundary.
$\bar{T}_{34}$, which is also large at $B = 0$, decreases to a prominent local minimum at $B \approx +0.002$, corresponding to the spectrum in Fig.~\ref{Tspec_B}(a)(ii).
At the minimum, a large portion of the wave coming in from lead $4$ is deflected into lead $1$, leading to a corresponding maximum in $\bar{T}_{14}$ [with $\bar{T}_{14}(B) = \bar{T}_{41}(-B) = \bar{T}_{32}(B)$, see Table~\ref{symmetry_table}], which is otherwise close to zero.
It is this drastic change at the intermediate field strength $B \approx +0.002$, different for each mean transmission component, that will serve as a key property to enable multiterminal transport control for the geometrical parameters used.
At even stronger field the waves from lead $4$ follow edge states directly into lead $3$, so that $\bar{T}_{34}$ increases again, while $\bar{T}_{14} = \bar{T}_{32}$ decreases.
For $B < 0$, $\bar{T}_{34}$ remains high over the local minimum of $\bar{T}_{21}$ around $B = -0.002$, before it too falls off for stronger fields.
$\bar{T}_{13}$, on the other hand, remains close to zero for all $B < 0$, increases with the field strength for $B > 0$, and finally follows $\bar{T}_{34}$ in the edge state regime, where the pathways $4 \rightarrow 3$ and $3 \rightarrow 1$ are almost equivalent along the boundary: indeed, the $T_{34}$ and $T_{13}$ spectra practically coincide in Fig.~\ref{Tspec_B}(a)(iv).

On all $\bar{T}_{ij}(B)$ curves, though more visible in $\bar{T}_{21}$ and $\bar{T}_{34}$, relatively small fluctuations in $B$ are superimposed, which can be regarded as generalized collective AB oscillations from interference between spatially extended leaking states:
the oscillations in ${T}_{ij}(B)$ at each $\kappa$ add up to a large-scale oscillation of the channel average.
The characteristics of $\bar{T}_{ij}$ remain qualitatively the same in the higher channels (with maximum $= n$), mapped onto a larger $B$-scale:
at higher energy larger field strength yields the same Larmor radius and similar variations as in $\bar{T}_{ij}(B)$.
From the above we see that, depending on $B$, overall transmission is favored from each lead to certain other leads and suppressed to the rest.
We will address this possibility for directed multiterminal transport in detail in Sec.~\ref{conductance}.

The modification of the $\bar{T}_{ij}(B)$ profiles by the perturbing disk is shown in Fig.~\ref{Tmean_Bvar}(i) to (v), where its radius $r$ is increased so that transmission between leads on opposite sides is suppressed, as previously described [see Figs. \ref{Tmean_rvar}(b) and \ref{Tspec_B}(b)].
Thus $\bar{T}_{21}(B)$, $\bar{T}_{34}(B)$ and $\bar{T}_{32}(B)$, although retaining their trends, gradually decrease to zero for any $B$ when the constrictions become narrower than the leads.
In contrast, $\bar{T}_{13}$ increases with $r$ at $B = 0$, as seen also in Fig.~\ref{Tmean_rvar}(b), and remains large at strong $B > 0$.
Also at strong $B < 0$, though, $\bar{T}_{13}$ increases with $r$, because the edge states (now clockwise deflected classical orbits) can guide the particle from lead $3$ onto the disk edge and then onto the elliptic boundary to the left of the disk, which it follows into lead $1$.
Interestingly, for large enough disk [as in Fig.~\ref{Tmean_Bvar}(v), where the constriction width is $w/3$] the 4-terminal billiard is effectively divided into two 2-terminal billiards for $n = 1$, so that transmission between leads on the same side of the disk becomes symmetric in $B$: 
$\bar{T}_{13}(B) = \bar{T}_{13}(-B)$ and $\bar{T}_{24}(B) = \bar{T}_{24}(-B)$, as a consequence of the sum rule Eq.~(\ref{sum_rule}) and the symmetry relation Eq.~(\ref{TRS}) for the special case of a system with two leads.\cite{Buttiker1988}
This 2-terminal symmetry is not present in the higher channels [$n = 2, 3$ in row (v)], where the smaller transversal wavelength enables transport through the constrictions.

Conclusively, the disk reduces the difference between the (independent) $\bar{T}_{ij}(B)$, thereby weakening the controllability of mutliterminal transmission.
Nevertheless, an appropriate disk-like blocking potential switches output from lead $2$ to lead $3$ (from lead $3$ to lead $2$) with input in lead $1$ (in lead $4$) at strong $B > 0$, which,
in the context of directed transport, constitutes an additional (electric) switching mechanism based on the geometry-independent behavior of edge states.

\subsection{Bent coupled wires\label{wires}}

\begin{figure}[t!]
    \begin{center}  
      \includegraphics[width=\columnwidth]{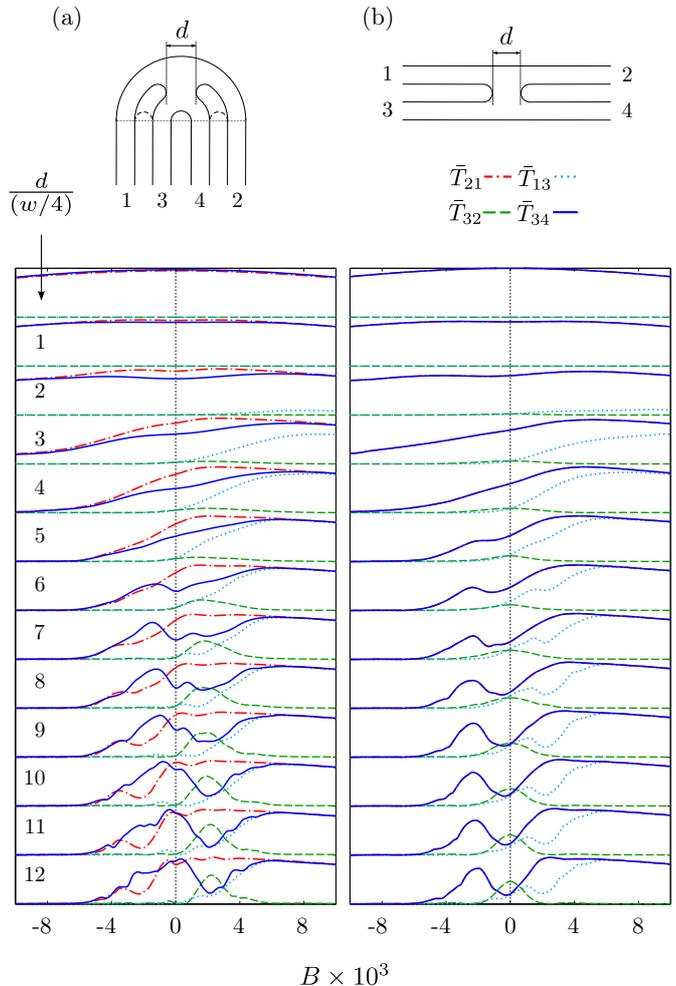}
     \end{center}
      \caption{(Color online) $\bar T_{ij}(B)$ for two parallel wires at distance equal to their common width $w = a/3.5$, coupled by a smooth opening of width $d$ and circular edges of radius $w/2$, (a) with the wires bent by an angle $\pi$ across the coupling and (b) in straight configuration (parallel and equidistant to the $x$-axis). In the top plots there is no opening, and then the opening width is increased in steps of $w/4$ from $d = w/4$ to $d = 3w$ (second from top to bottom, with transmission offsets decreasing by one), the latter yielding in (a) the semi-circular billiard with smooth lead openings.}
  \label{Tmean_B_wires}
\end{figure}

Magnetically induced directed transport in our setup was shown to rely on the convexity of its boundary, which enables highly transmittive pathways at low field strength by the formation of librator- and rotator-like modes.
These coexist in the semi-ellipse, so in order to separately investigate the role of boundary-localized modes in the overall transmission, and the impact of their interference with gradually upcoming bulk modes, we consider the following setup:
two circularly bent parallel quantum wires coupled through a smooth opening of adjustable width $d$ at the bend [see sketch in Fig.~\ref{Tmean_B_wires}(a)].
The $\bar{T}_{ij}(B)$ are plotted in Fig.~\ref{Tmean_B_wires}(a) without the opening (top) and for increasing opening width (second from top to bottom), and are compared to the case of straight coupled wires\cite{Hirayama1992,Bertoni2000} in Fig.~\ref{Tmean_B_wires}(b), where curvature is absent and additional $y$-symmetry is present.
With no opening the transmission in the two disconnected bent wires at $B = 0$ is almost perfect:
$\bar{T}_{21}$ and $\bar{T}_{34}$ depart from unity only due to narrow resonances caused by the curvature of the wires, which effectively induces an attractive potential.\cite{Nagaoka1992}
$\bar{T}_{21}$ is slightly smaller than $\bar{T}_{34}$, because the resonances for the longer bent part lie closer in $\kappa$.
Both slowly decrease at stronger fields, as the magnetoelectric subband threshold rises [like in Fig.~\ref{Tspec_B}(a)(iv)].

\begin{figure*}[t!]
    \begin{center}  
      \includegraphics[width=.9\textwidth]{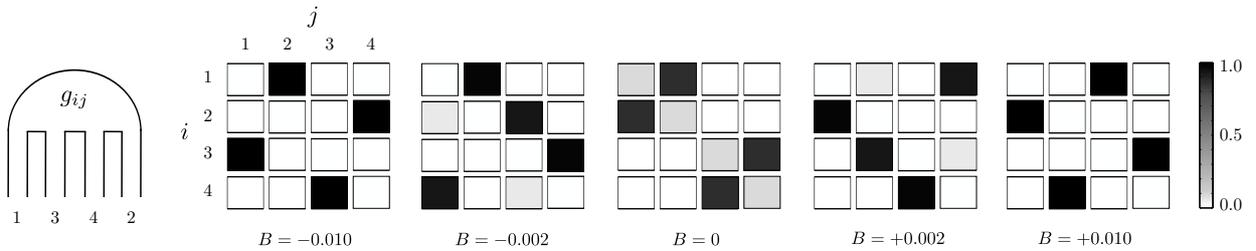}
     \end{center}
      \caption{Multiterminal conductance coefficients $g_{ij}$ at scaled Fermi energy $\kappa_F = w/\pi \cdot \sqrt{2 E_F} = 1.425$ and temperature $\Theta = 100~\rm{mK}$, for $B = 0$, $B = \pm 0.002$ and $B = \pm0.010$. In each $4 \times 4$ block, row $i$ labels the input and column $j$ the output current terminal.}
  \label{cond_B}
\end{figure*}

When the wires couple, the 2-terminal $B$-symmetry of the isolated wires is broken.
As the opening is widened, the rotating modes connecting the outer and inner leads increasingly interfere with states that extend into the opening, leading to enhanced back-scattering dips in the corresponding transmission coefficients and a following decrease in $\bar{T}_{21}$ and $\bar{T}_{34}$.
The role of rotator- and librator-like modes is here manifest in the inner-lead transmission as $d$ is varied at $B \approx 0$.
$\bar{T}_{34}$ is high for $d \approx 0$, where transport is dominated by rotator-like modes, decreases at intermediate $d$, where the rotators are destroyed by the opening in the inner convex boundary, and increases again for large $d$ ($\gtrsim 6w/4$), where the outer convex boundary focuses the wave function into librator-like modes.
At appropriate strength, the magnetic field favors transport between terminals on either side of opening by deflection of the particle orbits.
$\bar{T}_{32}$ increases to a local maximum at intermediate $B > 0$, which rises for larger $d$.
$\bar{T}_{13}$, on the other hand, rises significantly only at high $B > 0$ where the particle is guided by edge states.

For maximal opening $d = 3w$ the setup becomes the 4-terminal semi-circular billiard, now with smooth lead openings.
Controllability of output terminal, as described in Section \ref{magnetic}, then becomes optimal, and the $\bar T_{ij}(B)$ profiles are very similar to the ones in Fig.~\ref{Tmean_Bvar}(i).
This shows that the smoothness of the lead openings, as well as the small change in eccentricity and size, although clearly affecting the dynamics in the scattering system and thereby the detailed spectral features, leave the overall field dependence qualitatively unchanged.

In Fig.~\ref{Tmean_B_wires}(b) the strong field asymptotics of the $\bar T_{ij}(B)$ for the straight coupled wires coincide with those of the bent wires, since transport through edge states is rather affected by the topology, and not by the geometry, of the scatterer.
The first obvious difference here is that the additional reflection symmetry about the $x$-axis renders the components $\bar{T}_{21}$ and $\bar{T}_{34}$, whose difference was central in the discussion so far, identical.
That is, regardless of the field strength, high transmission between the outer leads can never be combined with low transmission between the inner leads, as it can for the bent wires.
Also due to spatial symmetry, $\bar{T}_{32}$ must now be symmetric in $B$, and its broad peak around $B = +0.002$ for the bent wires is shifted to $B = 0$.
As a result, crossed-lead transmission (between leads $2, 3$ or $1, 4$) can no longer be switched from high to low by inverting the field.
At weak fields, $\bar{T}_{34}(B)$ varies similarly for bent and straight wires for small $d \lesssim 6w/4$;
for large $d$ though, the straight wires, unlike the bent wires, yield low $\bar{T}_{34} = \bar{T}_{21}$ around $B = 0$, because of the absence of modes that are focused at the lead openings or guide the incoming wave along the boundary.
We conclude that the increased symmetry of the straight leads and the absence of the convex boundary reduce the possible combinations of magnetically induced transport directions between the terminals.

\subsection{Directed conductance\label{conductance}}

In the discussion so far we have utilized the channel-integrated mean transmission [Eq.~(\ref{mean_transmission})] as a tool to compactly describe the average response of the transmission to parameter changes.
The actual measurable conductance coefficients $g_{ij}$ connecting the current flowing inwards at terminal $i$ with the voltage differences to all other terminals $j$ are, in the linear response regime at temperature $\Theta$, given by the Landauer-B\"uttiker formula;\cite{Datta1995} in units of the (spin-degenerate) conductance quantum $2e^2/h$, 
\begin{eqnarray}
g_{ij}(\Theta;E_F) = \int_{-\infty}^{+\infty} \! T_{ij}(E)F(\Theta,E_F;E) \, dE
\label{LB}
\end{eqnarray}
where
\begin{eqnarray}
F(\Theta,E_F;E) & \equiv & -\frac{\partial f(E_F;E)}{\partial E} 
             \cr & = & \frac{1}{4k_B\Theta}~ \text{sech}^2\left(\frac{E - E_F}{2k_B\Theta}\right)
\label{Fth}
\end{eqnarray}
with $f(E_F;E)$ being the Fermi distribution for given Fermi energy $E_F$.
Thus, $g_{ij}$ essentially equals the thermally averaged transmission around the electron Fermi energy, 
with a width proportional to $\Theta$, and coincides with $T_{ij}$ at $\Theta = 0$.

The output controllability described in Sec.~\ref{magnetic} can be enhanced in terms of the conductance at low temperature and low Fermi energy, where transmission features are resolved in a smaller $\kappa$-range than the whole channel.
While the local maxima (minima) of $\bar{T}_{32}$ and $\bar{T}_{34}$ in Fig.~\ref{Tmean_Bvar}(i) suggest a maximal efficiency for magnetically directed transport of about 50\% in the unperturbed elliptic setup, adjusting temperature and Fermi energy appropriately can yield corresponding conductance maxima (minima) close to unity (zero) at these field strengths.
To provide an example, we choose the Fermi energy corresponding to $\kappa = \kappa_F = \sqrt{2E_F } \cdot w/\pi = 1.425$ [cf. transmission spectrum in Fig.~\ref{Tspec_B}(a)] and a temperature $\Theta = 100~\rm{mK}$.
The resulting conductance coefficients for the above determined optimal field strengths are shown in Fig.~\ref{cond_B} as grayscale cells for the individual $g_{ij}$ ordered like in Table~\ref{symmetry_table}.
The diagonal elements $g_{ii}$ following from Eq.~(\ref{LB}) do not contribute to calculated currents, but indicate the degree of reflectance for ballistic transport and show the depart from unity of the sum of conductances from or to the other ($\neq i$) terminals.
For the chosen parameters the conductance coefficients practically reach unity (black cells) for specific terminal combinations $i \rightarrow j$ and practically vanish for the rest (white cells), depending on the direction and strength of the field.
In particular, for each input lead $i$, the output can be switched selectively to any lead $j \neq i$ by appropriately tuning $B$.
This relies on the above discussed interplay of geometry and magnetic field effects.
For large $|B|$ (see $B =\pm 0.010$ blocks in Fig.~\ref{cond_B}) edge states form and conductance is determined by the topology of the boundary (in general directed edge state transport can also be implemented, but with finite potential barriers\cite{Beenakker1991,Sheng1997,Xie2007}).
At zero and intermediate $|B|$ the output is governed by interference of spatially extended scattering states leading, for this billiard, to overall high transmittivity between certain non-neighboring terminals, as seen in the $B = 0$ and $B = \pm 0.002$ blocks in Fig.~\ref{cond_B}.

This shows that the current can be efficiently directed from a given input to a selected output terminal.
Especially for zero and intermediate magnetic field strength it is even possible to construct a controllable cross-junction for the currents: 
choosing, e.g., leads $1$ and $3$ as input terminals and leads $2$ and $4$ as output, current is flowing from lead $1$ to $2$ and from lead $3$ to $4$ for $B = 0$. 
For $B = +0.002$ current is flowing mostly from lead $1$ to lead $4$ and from lead $3$ to lead $2$, thus exchanging the directed connections between the terminals with applied magnetic field. 

For higher $\Theta$ the $g_{ij}(\Theta;E_F)$ in Fig.~\ref{cond_B} are generally shifted to more intermediate values, because $F(\Theta,E_F;E)$ is broadened and additional spectral features are included in the integration around $E_F$ in Eq.~(\ref{LB}).
On the other hand, for very low $\Theta$ the transmission spectrum is highly resolved by $F(\Theta,E_F;E)$ and the $g_{ij}$ become sensitive to small changes in $E_F$.
These dependencies are not present in the channel means $\bar{T}^{(n)}_{ij}$, which therefore serve to estimate the field strength values suitable for directed transport in a given billiard;
with adjusted temperature and Fermi energy, the effect can then be optimized by slightly modifying the transmission features through fine-tuning of $B$, in order to controllably obtain maximal and minimal $g_{ij}$.

\section{Summary\label{summary}}

We have investigated the ballistic transport properties and low-temperature magnetoconductance of a 4-terminal semi-elliptic quantum billiard. 
Analyzing the strong dependency of the transmission on the magnetic field and geometrical parameters like eccentricity of the semi-ellipse and placement and width of the leads, we have shown how electrons can effectively and controllably be guided from one input lead to any other output lead, including the cross-switching of output from combined input.

At zero field strength electrons are guided between the two outer or the two inner leads by modes corresponding to classical rotator or librator trajectories, respectively.
The role of rotator and librator modes has been clarified by introducing a perturbing disk, thereby destroying modes selectively.
In this context we have shown that the transmission is robust under small perturbations and that the disk offers additional possibilities of transmission control.
The efficiency of the selective transmission has been optimized with respect to the geometric parameters. 
Here it turns out that the optimal position of the inner leads deviates from the classically expected one at the focal points of the ellipse and that an eccentricity of about 0.35 for equidistant terminals is optimal.

The application of the magnetic field allows to controllably change the conductance coefficients: at appropriate field strengths electrons coming from one input lead can be guided to any other output lead.
This results from the deflecting effect of the field at intermediate strength.
Further, by increasing the field strength to the edge state regime, conductance is mostly determined by the topology of the billiard. 
In this regime sharp edges cause mixing and interference of multiple edge states, leading to oscillations in the transmission spectra, while a smooth boundary guides the individual edge states without mixing.

To further examine the role of the elliptic boundary, i.e. of  the existence of rotator and librator modes, we have investigated the transmission through a pair of bent coupled quantum wires and compared it to the topological equivalent setup of two straight coupled wires.
The bent wires show a degree of control superior to the straight setup, but the semi-ellipse allows for the highest degree of conductance control.
While the channel-averaged transmission clearly shows the possibility of directed transport, very high switching efficiency is achieved at low but realistic temperatures and appropriate values of the Fermi energy.
The semi-ellipse qualifies as a magnetically controllable cross-junction for ballistic quantum transport.

\section*{ACKNOWLEDGEMENTS}

Financial support by the Deutsche Forschungsgemeinschaft under the contract Schm 885/11-1 is gratefully acknowledged.


\begin{thebibliography}{99}
\bibitem{note_classical}
It is pointed out that the subject of this work are the quantum effects of transport through the four-terminal semi-ellipse;
classical considerations, including the classical librator and rotator trajectories as well as the gyromagnetic radius of deflected orbits, provide a means of interpretation and comparison of limited validity, since we do not focus on the semiclassical regime.
\bibitem{softwall}
We here concentrate on the geometric focusing effect of the boundary, separately from the effect of its softness in experimental realizations; the latter quantitatively modifies, but does not eliminate, the geometrically induced transport characteristics, and will not be considered here.
\end{thebibliography}
\end{document}